\documentclass[aps,prb,twocolumn,superscriptaddress]{revtex4} 
\usepackage{graphicx}
\usepackage{color}
\usepackage{dcolumn}
\usepackage{bm}
\begin{document}

\title{Screening in YBa$_2$Cu$_3$O$_{7-\delta}$ at large wave vectors}

\author{Simo Huotari}
\affiliation{European Synchrotron Radiation Facility, B.P. 220, F-38043 Grenoble cedex, France}
\author{J. Aleksi Soininen}
\affiliation{Department of Physics, P.O.\ Box 64, FI-00014 University of Helsinki, Finland}
\author{Gy{\"o}rgy Vank{\'o}}
\affiliation{European Synchrotron Radiation Facility, B.P. 220, F-38043 Grenoble cedex, France}
\affiliation{KFKI Research Institute for Particle and Nuclear Physics, 
P.\ O.\ Box 49,  H-1525 Budapest, Hungary}
\author{Giulio Monaco}
\affiliation{European Synchrotron Radiation Facility, B.P. 220, F-38043 Grenoble cedex, France}
\author{Valerio Olevano}
\affiliation{Institut N{\'e}el, CNRS, B.P. 166, F-38042 Grenoble, France}

\date{\today}

\begin{abstract}
We present experimental inelastic x-ray scattering (IXS) and {\em ab
initio} time-dependent density-functional-theory (TDDFT) studies of
YBa$_2$Cu$_3$O$_{7-\delta}$.  The response of the low-lying Ba $5p$
and Y $4p$ core electrons is shown to interact strongly with the Cu
$3d$ and O $2p$ excitations, with important consequences on screening.
The agreement between IXS and TDDFT results is excellent, apart from a
new type of excitations, mainly related to loosely bound Ba electrons
and significantly affected by correlations. This points to correlation
mechanisms not fully described by TDDFT that might have a role in
giving rise to antiscreening.
\end{abstract}

\maketitle

%
%
%

\section{\label{sec:Introduction}Introduction}

Despite 20 years of research since the discovery of cuprates as
high-temperature superconductors (HTSC),\cite{HTSC} the nature of the
superconductivity and the pairing mechanism in these systems still
remain unknown.  Extensive studies of HTSC cuprates have been done
using the most varied techniques and on almost every observable
imaginable.  The dielectric function
$\varepsilon(\mathbf{r},\mathbf{r}',t-t')$ is in particular a key
quantity for superconductivity.  Its inverse $\varepsilon^{-1}$
measures the screening of the bare Coulomb repulsion and can directly
indicate reverse screening (anti-screening) spatial regions where the
interaction between two electrons is attractive rather than repulsive.
In these regions electrons pair up, giving rise to superconductivity.
The study of $\varepsilon^{-1}$ can thus unambiguously indicate such
domains and shed light to the pairing mechanism.

The dielectric function is probed often via its Fourier transform
$\varepsilon(\mathbf{q},\omega)$ as in reflectance and ellipsometry
spectroscopy, but usually only at negligible transferred momentum
$|\mathbf{q}|=q \to 0$.  This provides access only to long-range
screening ($|\mathbf{r}-\mathbf{r}'| \to \infty$).  In HTSC the
coherence length is of the order of few inter-atomic distances [e.g.\
1.5 nm in YBa$_2$Cu$_3$O$_{7-\delta}$ (YBCO) in the $ab$ plane
\cite{superconductivity}].  The antiscreening domains have to exist at
comparable or shorter distances.  In the large-$q$ range, which would
correspond to short-range correlations, experimental techniques are
scarce due to the high-kinetic-energy particles required as a probe.
As a consequence, this dynamic domain of
$\varepsilon^{-1}(\mathbf{q},\omega)$ for $q>0$ has been a no-man's
land with very few applicable experimental probes.  There are
basically only two techniques available: the electron energy-loss
(EELS) (Refs.\ \onlinecite{fink89} and \onlinecite{abajo09}) 
and inelastic x-ray scattering (IXS)
(Ref. \onlinecite{schulkebook}) spectroscopies. {  
Both, in principle, measure the
loss function, which is proportional to 
$-{\mathrm Im}\left[\varepsilon^{-1}(\mathbf{q},\omega)\right]$.}
EELS, however, is inherently
limited to relatively small values of $q$ due to an increasing
multiple scattering at large wave vectors.  IXS is left as the only
technique probing exchanged wave vectors that correspond to
inter-atomic distances, the typical range accessible to IXS being $q
\gtrsim$ 5 nm$^{-1}$. It can hence be a valuable technique to study
HTSC.

IXS studies in strongly-correlated materials have been so far limited
by the low count rates in high-$Z$ elements.  Thanks to the rapid
development in instrumentation, it is now becoming possible to apply
IXS even to HTSC cuprates.  Resonant IXS
(Refs. \onlinecite{ishii05,schulkebook,kotani01}) has already been used to study
such systems but these experiments do not probe the dielectric
function and should not be confused with non-resonant IXS which we use
here.  In particular on YBCO, EELS experiments
\cite{balzarotti88,chen88,yuan88,tarrio88,romberg90} were performed
soon after the discovery of its superconductivity, but there are no
existing IXS experiments. This work is an attempt to close the gap in
our accessible kinematic probing range of excitation energies and
momenta.

\begin{figure*}
\includegraphics[width=0.8\linewidth]{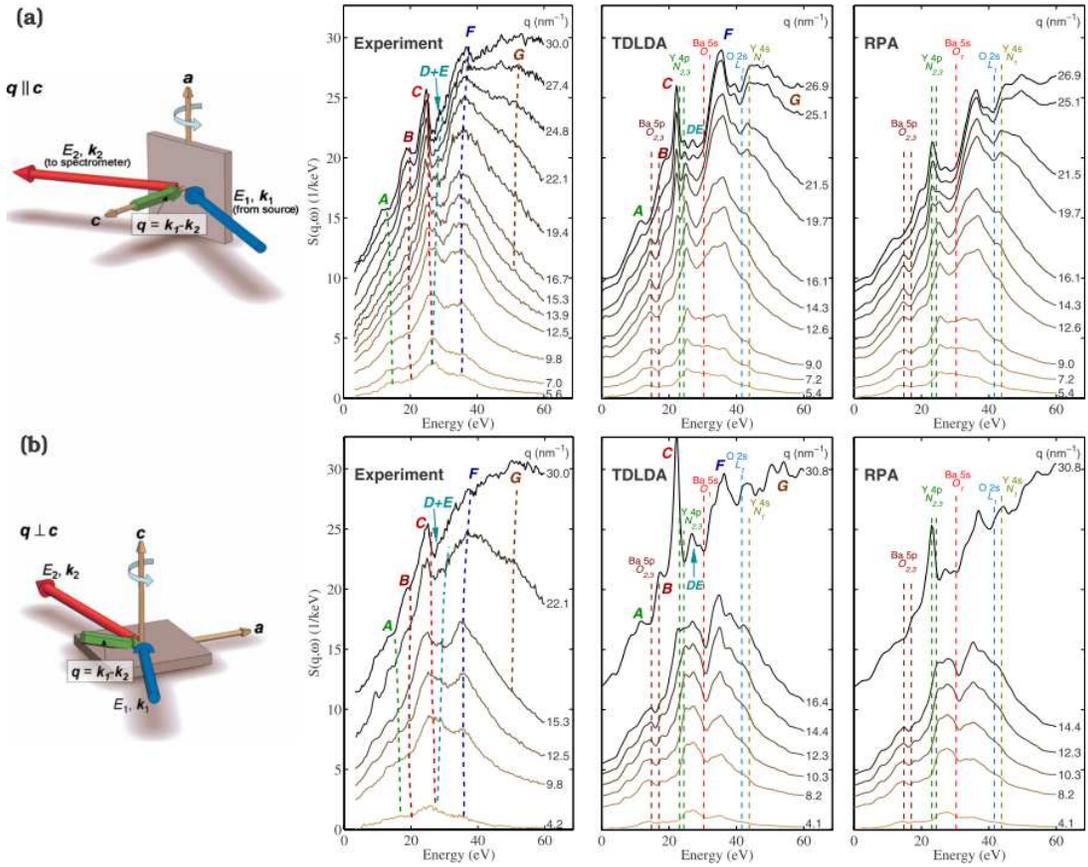}
\caption{\label{figu:exp_curves} 
{ Experimental and calculated energy-loss functions of YBCO for
(a) $\mathbf{q}~||~\mathbf{c}$ and (b) $\mathbf{q}\perp\mathbf{c}$.}
The most prominent spectral features are
labeled $A$--$G$ and highlighted with dashed lines in the experimental
plots.  The nominal core-electron excitation energies are marked with
dashed lines and labeled in the theory plots. For clarity, the spectra
are shifted vertically from each other by an amount proportional 
to the corresponding momentum transfer value. { Note that 
the experimental $S(\mathbf{q},\omega)$ are 
area-normalized to the TDLDA theory as described in the text.}
%
The experimental geometries for the two directions of $\mathbf{q}$ are 
shown on the left. 
}
\end{figure*}

The dominant theoretical picture on HTSC relies on the Hubbard model
\cite{Hubbard-HTSC} and invokes a strong correlation mechanism to
explain the pairing and the superconductivity in
cuprates.\cite{Hubbard-review} This model accounts only for the last
unpaired electron of a copper atom and disregards all the complicated
chemical, atomic and electronic structures of HTSC cuprates. {
They} are supposed only to renormalize the two adjustable
parameters of the model.  { Today}, a good framework to search
for a solution to the Hubbard model is the dynamical mean-field theory
(DMFT) (Ref.\ \onlinecite{georges}) and more recently cellular DMFT which
introduces a $\mathbf{k}$ dependence necessary to address the evident
anisotropy in HTSC. \cite{ybco-cdmft,htsc-cdmft} However, systematic
improvements within this approach are difficult without introducing
new adjustable parameters.

It has been recently demonstrated that \textit{ab initio} theories can
be applied to HTSC.  Density-functional theory (DFT) has shown
\cite{ybcolda,ybcopressure} its ability to reproduce the ground-state
atomic structure of YBCO within its typical error $<3\%$.  However,
DFT does not describe excited states.  On the other hand,
time-dependent density-functional theory (TDDFT)
(Refs.\ \onlinecite{runge-gross-84} and \onlinecite{gross-kohn-85}) 
is in principle an {\em exact}
theory to describe neutral excitations and the dielectric function.
In practice, since the exact exchange-correlation functional of TDDFT
is not available at the moment, approximations such as the adiabatic
local-density approximation (TDLDA) (Ref.\ \onlinecite{zangwill-soven}) are
required.  { The latter has shown to provide loss-function
spectra for ordinary
semiconductors and insulators, e.g. silicon, in good agreement with
both EELS (Ref.\ \onlinecite{tddft-eels}) and IXS 
(Ref.\ \onlinecite{tddft-ixss}) experiments.  }
Recent works have shown that TDLDA provides results in excellent
agreement { also} with the exact solution of a one-dimensional
Hubbard model, reproducing spin and charge collective excitations in a
Luttinger liquid. \cite{tddftluttinger,tddftaldaluttinger} TDLDA can
hence be a good formalism even in strongly correlated systems.
However, so far there are no TDDFT studies of the dielectric function
in cuprates.
 
In this work we present an investigation of the inverse dielectric
function $\varepsilon^{-1}(\mathbf{q},\omega)$ on a prototypical 
high-temperature superconductor cuprate, YBa$_2$Cu$_3$O$_{7-\delta}$
($\delta=0.07$) combining IXS experiments with TDDFT calculations. IXS
probes the dynamic structure factor $S(\mathbf{q},\omega)$, which is
related to the dielectric function via 
\begin{equation}
S(\mathbf{q},\omega) = -(\hbar
q^2)/(4\pi^2e^2n) ~ \mathrm{Im}~\varepsilon^{-1}(\mathbf{q},\omega)
\end{equation}
$n$ being the electron density. We study
$S(\mathbf{q}~||~\mathbf{c},\omega)$ 
{ and $S(\mathbf{q}\perp\mathbf{c},\omega)$} 
for energies of 5--60 eV and $q$
between 5.6--30.0 nm$^{-1}$.  The comparison of experimental results
with TDDFT calculations provides an important benchmark for studying
the electronic response of cuprates in the whole physically relevant
momentum-transfer range.  In particular, the large energy range
studied here together with calculations facilitate the identification
of the role of yttrium and barium in the valence electron dynamics of
YBCO.

%
%

\section{\label{sec:Experiment}Experiment}

The non-resonant IXS measurements of the loss function of YBCO were
done at the beamline ID16 (Refs.\ \onlinecite{verbeni09} and \onlinecite{huotari05}) of
the European Synchrotron Radiation Facility (ESRF).  The measurements
were performed at room temperature using monochromatic x rays with
energies of 7.9--8.0~keV.  The spectrometer was based on a spherically
bent Si(444) analyzer crystal in the Rowland-circle geometry. The
bending radius of the crystal was 1~m and the Bragg angle was fixed to
89$^\circ$. The spectrometer observed the intensity of scattered
photons with a fixed energy, and the energy transfer was tuned by
changing the incident-photon energy. { The sample was in a
shape of a plaquette with a size of 5$\times$5$\times$0.1 mm$^3$, with
the polished main face 
oriented perpendicular to the $\mathbf{c}$ axis and the edges
along the $\mathbf{a}$ and $\mathbf{b}$ axes. The
sample orientation was analyzed using a Laue photograph. The 
superconducting transition of the sample takes place at $T_C=94$~K. 
The measurements were performed with $\mathbf{q}~||~\mathbf{c}$ and with 
$\mathbf{q}\perp\mathbf{c}$.  The measurement geometries
in the two cases are shown in Fig.\ \ref{figu:exp_curves}. While the
sample is easy to align unambiguously for 
the measurements with $\mathbf{q}~||~\mathbf{c}$, the measurements with
$\mathbf{q}\perp\mathbf{c}$ were slightly more difficult with the sample
in question. First of all, the usual
twinning of a YBCO was present in the sample, giving a small amount 
of uncertainty to the exact directions of the $\mathbf{a}$ and $\mathbf{b}$ 
axes. While the measurements were performed as close to $\mathbf{q}~||~\mathbf{a}$ as possible, for this reason we refer to these results rather as 
$\mathbf{q}\perp\mathbf{c}$. Furthermore, the
measurement in the $ab$ plane was done in a near-grazing incidence
and near-grazing exit geometry. 
The incidence and exit angles were thus small but finite, giving a small
$\mathbf{c}$ component to the momentum-transfer vector as shown in 
Fig.\ \ref{figu:exp_curves}(b). However, this component
was kept always smaller than the $\mathbf{q}$ resolution and thus is
expected to have a negligible effect on the results.

All spectra at each
fixed $\mathbf{q}$ were measured several times in the same conditions
and after a normalization to the incident-beam intensity were found to
be identical
%
within statistical accuracy. 
The resulting spectra were then averaged in order to
increase
%
the 
statistical accuracy. This procedure minimizes the
possibility for any experimental instabilities during the
measurements.  The resulting spectra, which at this stage are
scattering intensities as a function of energy transfer $\omega$ and
momentum transfer $\mathbf{q}$ were corrected for sample
self-absorption and linear background.  We refer to the $\mathbf{Q}=0$
as the zone center, and thus momentum transfer values given below are
absolute.  } The energy resolution of the experiment was 1.0~eV and
the momentum-transfer resolution 1.5 nm$^{-1}$. The zero-energy-loss
quasielastic line was subtracted by using an exponential fit to its
positive-energy-transfer tail above 2.5 eV. This procedure yields
reliable spectra above $\sim3$~eV. The experimental spectra were
normalized to the same area with the theoretical TDLDA spectra.
{ Since 
%
in the current computational approach~\cite{dp-web}
the theoretical results are limited to values 
$q=\frac{2\pi}{nx}$, 
where $n$ is an integer and $x$ the length of the direct-space
lattice vector, it was not
necessarily possible to perform the calculation of $S(\mathbf{q},\omega)$ 
exactly at all measured values of $q$. For the determination of the
area-normalization factor, we first determined numerically 
the theoretical behavior of the integrated area 
$S_\mathrm{int}^\mathrm{theory}(q) = \int_0^{60 \mathrm{eV}} S(q,\omega) \mathrm{d} \omega$
and interpolated the resulting curve to the values of $q$ which were actually
measured. A normalization to this area-integrated value yields finally
the experimental $S(\mathbf{q},\omega)$.
}

%
%

\section{\label{sec:Theory}Theory}

Our computational starting point was a standard ground-state DFT LDA
calculation (as in Ref.~\onlinecite{ybcolda}) of the total energy and
the electronic density in YBCO using the \textsc{ABINIT} code.
\cite{abinit-web} A plane-wave basis set was used with a converged 160
Ry kinetic-energy cutoff and periodic boundary conditions.  As turns
out later to be very important, we included the low-lying-core states
Ba $5s5p$, Y $4s4p$, Cu $3s3p$, and O $2s$ in the valence using
Hartwigsen-Goedecker-Hutter pseudopotentials.  Employing the
calculated DFT Kohn-Sham energies and wave functions we carried out a
TDDFT calculation of the polarizability and the dielectric function
using the \textsc{DP} code, \cite{dp-web} { using both TDLDA
and random phase approximation (RPA). The latter neglects
correlation effects { in the dielectric response of the system}, 
and the comparison of TDLDA and RPA might give a
hint where correlation effects play an important role. } The
independent-particle polarizability was constructed summing over 350
Kohn-Sham bands calculated at a Monkhorst-Pack $8\times8\times3$
$k$-point grid shifted by ($\frac{1}{2}$ $\frac{1}{2}$ $\frac{1}{2}$).
The dimensions of polarizabilities and { longitudinal}
dielectric functions as matrices in the reciprocal space were
347$\times$347. The inversion of the polarizability matrix produces
so-called local field effects where different excitation channels
mix. This effect, usually neglected in the localized description, is
well known from earlier first-principles studies on, e.g., transition
metals. \cite{gurtubay05} It should be emphasized that we found these
local-field { effects to be crucial} to obtain a good
agreement between experiment and theory and should definitely not be
neglected in this system.  We used experimental lattice parameters,
and we verified that the relaxed theoretical atomic structure
\cite{ybcolda} produces only minor changes to the spectra.  All
theoretical data shown here were convoluted with the experimental
resolution function.  
%
%

%
%

\section{\label{sec:Results}Results and discussion}

{ 
The measured and calculated spectra are shown in Fig.\ \ref{figu:exp_curves}.}
The agreement between the experimental and theoretical results is
remarkably good. This is a manifestation that the TDDFT is an
extremely useful tool in understanding the physics of even this
canonical strongly correlated system. In fact, all relevant physics of
the electronic excitations studied here are captured by the TDDFT.
The energy range of the investigated excitations covers the collective
oscillations of the valence electrons (plasmons), consisting mostly of
the hybridized Cu $3d$ and O $2p$, and the excitations of shallow core
electrons of Y and Ba. Because the latter interact strongly with the
valence electrons due to the important overlap in energy, we call them
{\em semicore states}.  The literature values
\cite{fuggle80,cardona78} for their atomic excitation energies are
14.8 eV (Ba $5p_{3/2}$), 17.0 eV (Ba $5p_{1/2}$), 23.1 eV (Y
$4p_{3/2}$), 24.4 eV (Y $4p_{1/2}$), 30.3 eV (Ba $5s$), 41.6 eV (O
$2s$) and 43.8 eV (Y $4s$).  These are marked in the Fig.\
\ref{figu:exp_curves} with dashed lines. Assuming that these electrons
indeed are strongly bound, they can only be excited above these
energies. 
{ Measuring the spectra at these core-electron
excitation thresholds to unoccupied states above the Fermi level 
bears 
resemblance 
to soft x-ray absorption spectra of those core
electrons. This scattering process is called x-ray Raman scattering,
and has been used successfully to measure
soft x-ray absorption spectra of low-lying core electrons edge even in
high pressure environments, accessible by the high-energy x-rays used
in the experiment. \cite{pylkkanen10,lee08} }

The IXS spectra in this regime are similar to those that could be
studied by EELS except for the important possibility to access large
momentum transfers (up to 25 nm$^{-1}$ in this study but in principle
unlimited).  For example, an early EELS study of YBCO has revealed
spectra equivalent to those reported here at the lowest $q$
(Ref.\ \onlinecite{romberg90}). 
{
The anisotropy between the $ab$ plane and the $c$-axis 
results is surprisingly small in the experiment. Meanwhile, 
the theory (both within TDLDA and RPA) would predict a slightly
larger anisotropy, \cite{tddft-anisotropy} 
at least between the calculated $\mathbf{a}$ and 
$\mathbf{c}$ directions. 
}

\begin{figure}
\includegraphics[width=\linewidth]{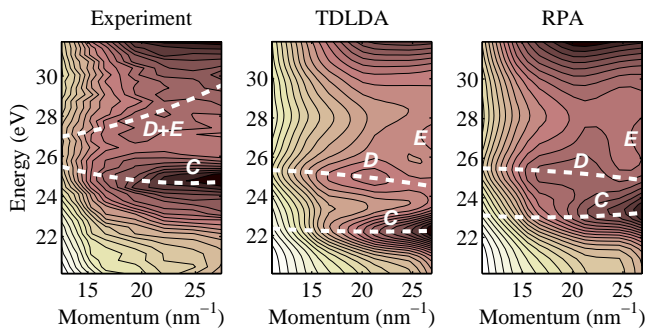}
\caption{\label{figu:contourf} Zoom-up to the
region of 20--32~eV energy transfer to show the Ba $5p$ + Y $4p$
related dispersing excitations $C$--$E$ in experiment, TDLDA and RPA.
Their dispersion are marked with dashed lines as a guide for eyes.}
\end{figure}

The spectral features found here are labeled with
letters $A$--$G$ in Fig.\ \ref{figu:exp_curves}. The features $A$ and $B$
correspond to the excitation of the Ba $5p$ electrons (the Ba
$O_{2,3}$ absorption edges).  The features $C$--$E$ are rather
intriguing.  
{ The relevant $S(q,\omega)$ range for the case of 
$\mathbf{q}~||~\mathbf{c}$ is showed zoomed-up in
the Fig.\ \ref{figu:contourf}, in both TDLDA and RPA together with the
experiment.}  It is in this region where the differences of TDLDA and
RPA are the most pronounced. In RPA the peaks $D$ and $E$ are broad
and weak, but appear better defined in TDLDA.  On the other hand,
experimentally there are only two excitations, which we label
tentatively as $C$ and $D+E$ since it seems that the peaks $D+E$ have
merged together within the experimental accuracy.  All these peaks
appear at higher energies in the experiment than in the theory.  The 
energies of the excitations $C$--$E$ disperse with momentum transfer,
for the excitations $D+E$ most strongly in the experimental result.
As a guide for the eyes, their dispersion is marked with dashed lines
in Fig.\ \ref{figu:exp_curves}.  In the experiment $C$ and $D+E$ seem to
coalesce into a single peak at low momentum transfer.  To observe
these excitations separately, an access to large momentum transfers is
essential, in particular since it seems that the peak $E$ starts
gaining considerable spectral weight only for $q>20$ nm$^{-1}$.  These
peaks are reproduced by the theory in both approximations but a clear
discrepancy between experiment and theory remains.  In particular in
the experiment the excitation $D+E$ seems to follow a quadratic
dispersion typical for a plasmon, a trend not reproduced by theory
even if TDLDA reproduces the excitations better.  {\em This could be
one of the key differences where a treatment of correlations beyond
TDLDA is necessary.}\cite{tddft-longrange,tddft-dynamical}
These excitations are sensitive to correlation,
as suggested by the evident differences between the two
exchange-correlation approximations.  It would seem natural to assign
them to Y $4p$ which has an excitation threshold in the same energy.
However, surprisingly, by occupation analysis we have found
them to originate mostly from Ba $5p$ instead.  As the wave functions
of these excited states overlap with those of Y $4p$, the interaction
of Ba $5p$ and Y $4p$ excitations seems to be strong.  Finally the
features $F$ and $G$ are related to the Ba $5s$ and Y $4s$ + O $2s$
excitations, respectively.

\begin{figure}
\includegraphics[width=\linewidth]{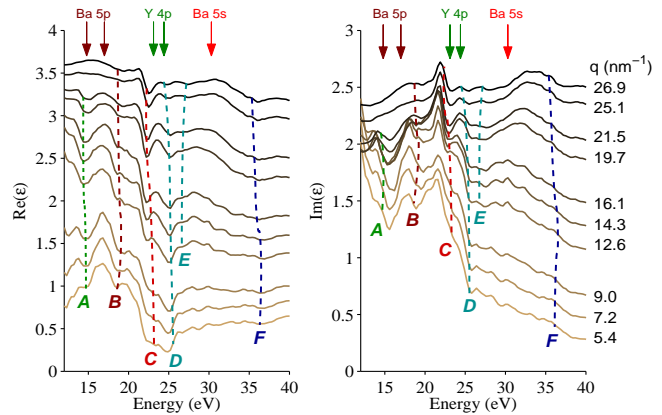}
  \caption{\label{figu:dielefun}The real part (left) and 
imaginary part (right) of the dielectric function as calculated within the 
TDLDA for the $\mathbf{q}~||~\mathbf{c}$ direction. 
The excitation dispersions are labeled as in Fig.\ \ref{figu:exp_curves}
and nominal core-excitation thresholds are annotated on top.}
\end{figure}

The dielectric function as calculated within the TDLDA is shown in
Fig.\ \ref{figu:dielefun} for the $\mathbf{q}~||~\mathbf{c}$ direction.
 Any single-particle excitation, associated to
a transition of a single electron from an occupied to an empty
state, should be apparent as a peak in Im($\varepsilon$).
The position of such an excitation corresponds to the
transition energy
between the two states of the band-like picture,
eventually renormalized by electron-electron
interaction effects
(single-quasiparticle\cite{quasipart-effects} excitations)
and electron-hole 
(excitonic\cite{excitonic-effects,excitonic-effects-rep}) effects.

On the other hand, collective plasmon excitations would
manifest themselves as zero-crossings of the real part of the
dielectric function Re($\varepsilon$), when the imaginary part
Im($\varepsilon$) is small.
{ These frequencies are in correspondence with
the self-sustained resonant modes of the system, associated to
collective oscillations of the electron plasma.}
When the plasmon is strongly damped (usually due to an interaction
with single-particle excitations), the zero-crossing disappears but a
relict may remain as a minimum of Re($\varepsilon$).  Here we find no
zero-crossings of Re($\varepsilon$) in agreement with earlier studies,
\cite{romberg90} implying that the plasmons in this system are heavily
damped.  While the peaks $A$--$C$ are produced by dips in the real
part at low $q$, they change their character into a peak in the
imaginary part at highest $q$ studied here, showing a transition from
a highly-damped plasmon to a single-particle excitation as $q$ is
increased. The peak $D$ has the clearest damped-plasmon-type character
with the largest minimum in Re($\varepsilon$) at low $q$.  While much
of the spectral weight is due to the Cu $3d$ and O $2p$, most of the
fine structure is in fact due to the Y $4p$ and Ba $5p$ electrons.  By
occupation analysis we have identified the peaks $A$--$D$ to originate
from Ba $5p$, but especially at the largest momentum transfers the Y
$4p$ single-particle core excitations overlap with the excitation $D$.
In other words the Y $4p$ and Ba $5p$ core excitations interact
strongly with the Cu $3d$ + O $2p$ valence excitations.  For the peak
$E$ there is a weak peak in Im($\varepsilon$) at the corresponding
energy and momentum, suggesting a single-electron type excitation for
this peak, mostly Y $4p$ character.

A surprising aspect in the charge response of YBCO in this energy
range is that it seems to fall outside the above-described normal
separation between collective and single-particle excitations, and
rather has a peculiar mixed character. For instance, we found that at
low $q$ the changes in Re($\varepsilon$) due to Ba $5p$, often
considered as a core state, seem to be indispensable for obtaining a
quantitative agreement between the theory and the experiment. The Ba
$5p$ excitations introduce new minima in Re($\varepsilon$) around
25~eV reducing its value by approximately 50\%, giving a large
contribution to the collective $C$-$D$ excitations at low $q$.
Furthermore, below 20 eV the Ba $5p$ electrons increase the value of
Re($\varepsilon$) and decrease the spectral weight in this energy
range in the final loss function. This is even true below 10 eV where
the increase in Re($\varepsilon$) is approximately 10\%.  However, at
$q \gtrsim 20$nm$^{-1}$ their response is seen as peaks in
Im($\varepsilon$), indicating in turn typical core-electron
excitations.  On the contrary, the Y $4p$ excitations have mostly a
single-particle character at all values of $q$ studied here.
Nevertheless, they have a significant role in determining the final
shape of the loss function.
     
The results shown here are a clear proof that the valence dynamics of
YBCO cannot be described in detail by considering only the Cu $3d$ and
O $2p$, but at least the Ba $5p$ and Y $4p$ have to be taken into
account.  Besides hybridizing, they couple to the Cu $3d$ + O $2p$
through the local field effect and they may be important also at much
lower energies than the ones investigated here.

{ { All real electronic systems, and even the jellium model,}
exhibit damping of
the collective excitations at increased wave vectors --- a result well
known from Landau's theory (see for instance the discussion 
in Refs.\ \onlinecite{galambosi05} and \onlinecite{huotari09} 
and references therein). 
%
One of the reasons for the { plasmon damping is the} coupling 
to single-particle electron-hole pair excitations.
In simple metals the effect of 
single-particle excitations on the plasmon linewidth has been studied using,
e.g., nearly-free-electron approach\cite{sturm81} as well as more 
recently using an  \textit{ab initio} method.~\cite{eguiluz00} 
In all cases where single-particle excitation channels are
strong in the range of plasmon energies, 
{ the damping will be more severe than in the jellium model.}
This is exactly the reason why a single-band theory { cannot
describe this situation}: all electronic excitation pathways, which
depend very much on the { full} band structure,
have to be taken into account equally. 
In the cuprates and other so-called strongly correlated
systems this may not be a trivial task.
{ Even excitations
of semicore electrons can play an important role in the damping
of collective excitations, as we have shown in YBCO.
On the other hand, it is not yet clear whether the band-like
paradigm still holds for the highest occupied valence 
levels in these systems.}
This was exactly the
motivation for the current work. 
We have shown here that {\em ab initio}
TDDFT can describe { a complicated system like YBCO } and 
increase our understanding of the interplay between 
semicore and valence electrons.}
%


\section{\label{sec:Conclusions}Conclusions}

In conclusion, we have presented a combined experimental and
theoretical study of the dielectric function of YBCO for large wave
vectors, probing small interparticle distances.  This previously
unexplored area of studies reveals new interactions between Cu $3d$
and O $2p$ valence excitations with the weakly bound Ba $5p$ and Y
$4p$ electrons in the dynamic response and screening. Especially the
role of Ba $5p$ is found to be of mixed single-particle and
damped-collective type giving rise to a novel excitation, observable
only at large wave vectors, suggesting its importance in screening at
short distances.  We have shown that the valence dynamics interacts
strongly with the low-lying semicore states.  We have answered as well
the question of how well we are able to model a HTSC cuprate system
with many-particle interactions of the valence electrons that
invalidate the band structure picture.  In fact, against all
expectations, {\em ab initio} TDDFT proves to be a valid and powerful
theory to describe a HTSC cuprate and gives results which agree with
the experiment very well.  Finally, it is possibly in the remaining
differences between the experiment and theory presented in this work
where the effects of strong correlations, and perhaps even the origin
of the Cooper pairing, could be looked for.

Beamtime was granted by ESRF and computing time by Ciment via
NanoSTAR.  J.A.S.\ acknowledges the Magnus Ehrnrooth foundations
support for this research and Academy of Finland (Contracts No.\
1110571 and 1127462).  G.V.\ acknowledges financial support from the
Hungarian Research Fund (OTKA, contract No. K 72597) and from the
Bolyai Janes program of the Hungarian Academy of Sciences.  
The authors would like to thank Roberto Verbeni
and Christian Henriquet for their support.


\begin{thebibliography}{45}
\expandafter\ifx\csname natexlab\endcsname\relax\def\natexlab#1{#1}\fi
\expandafter\ifx\csname bibnamefont\endcsname\relax
  \def\bibnamefont#1{#1}\fi
\expandafter\ifx\csname bibfnamefont\endcsname\relax
  \def\bibfnamefont#1{#1}\fi
\expandafter\ifx\csname citenamefont\endcsname\relax
  \def\citenamefont#1{#1}\fi
\expandafter\ifx\csname url\endcsname\relax
  \def\url#1{\texttt{#1}}\fi
\expandafter\ifx\csname urlprefix\endcsname\relax\def\urlprefix{URL }\fi
\providecommand{\bibinfo}[2]{#2}
\providecommand{\eprint}[2][]{\url{#2}}

\bibitem[{\citenamefont{Bednorz and M{\"u}ller}(1986)}]{HTSC}
\bibinfo{author}{\bibfnamefont{J.~G.} \bibnamefont{Bednorz}} \bibnamefont{and}
  \bibinfo{author}{\bibfnamefont{K.~A.} \bibnamefont{M{\"u}ller}},
  \bibinfo{journal}{Z. Phys. B} \textbf{\bibinfo{volume}{64}},
  \bibinfo{pages}{189} (\bibinfo{year}{1986}).

\bibitem[{\citenamefont{Buckel and Kleiner}(2004)}]{superconductivity}
\bibinfo{author}{\bibfnamefont{W.}~\bibnamefont{Buckel}} \bibnamefont{and}
  \bibinfo{author}{\bibfnamefont{R.}~\bibnamefont{Kleiner}},
  \emph{\bibinfo{title}{Superconductivity}} (\bibinfo{publisher}{Wiley},
  \bibinfo{address}{Weinheim}, \bibinfo{year}{2004}).

\bibitem[{\citenamefont{Fink}(1989)}]{fink89}
\bibinfo{author}{\bibfnamefont{J.}~\bibnamefont{Fink}}, \bibinfo{journal}{Adv.
  Electron. Electron Phys.} \textbf{\bibinfo{volume}{75}}, \bibinfo{pages}{121}
  (\bibinfo{year}{1989}).

\bibitem[{\citenamefont{de~Abajo}(2010)}]{abajo09}
\bibinfo{author}{\bibfnamefont{F.~J.~Garc{\'i}a} \bibnamefont{de~Abajo}},
  \bibinfo{journal}{Rev. Mod. Phys.} \textbf{\bibinfo{volume}{82}},
  \bibinfo{pages}{209} (\bibinfo{year}{2010}).

\bibitem[{\citenamefont{Sch{\"u}lke}(2007)}]{schulkebook}
\bibinfo{author}{\bibfnamefont{W.}~\bibnamefont{Sch{\"u}lke}},
  \emph{\bibinfo{title}{Electron Dynamics by Inelastic X-Ray Scattering}}
  (\bibinfo{publisher}{Oxford University Press, Oxford}, \bibinfo{year}{2007}).

\bibitem[{\citenamefont{Ishii et~al.}(2005)\citenamefont{Ishii, Tsutsui, Endoh,
  Tohyama, Kuzushita, Inami, Ohwada, Maekawa, Masui, Tajima et~al.}}]{ishii05}
\bibinfo{author}{\bibfnamefont{K.}~\bibnamefont{Ishii}},
  \bibinfo{author}{\bibfnamefont{K.}~\bibnamefont{Tsutsui}},
  \bibinfo{author}{\bibfnamefont{Y.}~\bibnamefont{Endoh}},
  \bibinfo{author}{\bibfnamefont{T.}~\bibnamefont{Tohyama}},
  \bibinfo{author}{\bibfnamefont{K.}~\bibnamefont{Kuzushita}},
  \bibinfo{author}{\bibfnamefont{T.}~\bibnamefont{Inami}},
  \bibinfo{author}{\bibfnamefont{K.}~\bibnamefont{Ohwada}},
  \bibinfo{author}{\bibfnamefont{S.}~\bibnamefont{Maekawa}},
  \bibinfo{author}{\bibfnamefont{T.}~\bibnamefont{Masui}},
  \bibinfo{author}{\bibfnamefont{S.}~\bibnamefont{Tajima}},
  \bibnamefont{et~al.}, \bibinfo{journal}{Phys. Rev. Lett.}
  \textbf{\bibinfo{volume}{94}}, \bibinfo{pages}{187002}
  (\bibinfo{year}{2005}).

\bibitem[{\citenamefont{Kotani and Shin}(2001)}]{kotani01}
\bibinfo{author}{\bibfnamefont{A.}~\bibnamefont{Kotani}} \bibnamefont{and}
  \bibinfo{author}{\bibfnamefont{S.}~\bibnamefont{Shin}},
  \bibinfo{journal}{Rev. Mod. Phys} \textbf{\bibinfo{volume}{73}},
  \bibinfo{pages}{203} (\bibinfo{year}{2001}).

\bibitem[{\citenamefont{Balzarotti et~al.}(1988)\citenamefont{Balzarotti, {M.
  De Crescenzi}, Motta, Patella, and Sgarlata}}]{balzarotti88}
\bibinfo{author}{\bibfnamefont{A.}~\bibnamefont{Balzarotti}},
  \bibinfo{author}{\bibnamefont{{M. De Crescenzi}}},
  \bibinfo{author}{\bibfnamefont{N.}~\bibnamefont{Motta}},
  \bibinfo{author}{\bibfnamefont{F.}~\bibnamefont{Patella}}, \bibnamefont{and}
  \bibinfo{author}{\bibfnamefont{A.}~\bibnamefont{Sgarlata}},
  \bibinfo{journal}{Solid State Comm.} \textbf{\bibinfo{volume}{68}},
  \bibinfo{pages}{381} (\bibinfo{year}{1988}).

\bibitem[{\citenamefont{Chen et~al.}(1988)\citenamefont{Chen, Schneemeyer,
  Liou, Hong, Kwo, Chen, and Waszczak}}]{chen88}
\bibinfo{author}{\bibfnamefont{C.~H.} \bibnamefont{Chen}},
  \bibinfo{author}{\bibfnamefont{L.~F.} \bibnamefont{Schneemeyer}},
  \bibinfo{author}{\bibfnamefont{S.~H.} \bibnamefont{Liou}},
  \bibinfo{author}{\bibfnamefont{M.}~\bibnamefont{Hong}},
  \bibinfo{author}{\bibfnamefont{J.}~\bibnamefont{Kwo}},
  \bibinfo{author}{\bibfnamefont{H.~S.} \bibnamefont{Chen}}, \bibnamefont{and}
  \bibinfo{author}{\bibfnamefont{J.~V.} \bibnamefont{Waszczak}},
  \bibinfo{journal}{Phys. Rev. B} \textbf{\bibinfo{volume}{37}},
  \bibinfo{pages}{9780} (\bibinfo{year}{1988}).

\bibitem[{\citenamefont{Yuan et~al.}(1988)\citenamefont{Yuan, Brown, and
  Liang}}]{yuan88}
\bibinfo{author}{\bibfnamefont{J.}~\bibnamefont{Yuan}},
  \bibinfo{author}{\bibfnamefont{L.~M.} \bibnamefont{Brown}}, \bibnamefont{and}
  \bibinfo{author}{\bibfnamefont{W.~Y.} \bibnamefont{Liang}},
  \bibinfo{journal}{J. Phys. C: Solid State Phys.}
  \textbf{\bibinfo{volume}{21}}, \bibinfo{pages}{517} (\bibinfo{year}{1988}).

\bibitem[{\citenamefont{Tarrio and Schnatterly}(1988)}]{tarrio88}
\bibinfo{author}{\bibfnamefont{C.}~\bibnamefont{Tarrio}} \bibnamefont{and}
  \bibinfo{author}{\bibfnamefont{S.~E.} \bibnamefont{Schnatterly}},
  \bibinfo{journal}{Phys. Rev. B} \textbf{\bibinfo{volume}{38}},
  \bibinfo{pages}{921} (\bibinfo{year}{1988}).

\bibitem[{\citenamefont{Romberg et~al.}(1990)\citenamefont{Romberg, N{\"u}cker,
  Fink, Wolf, Xi, Koch, Geserich, Durrler, Assmus, and
  Gegenheimer}}]{romberg90}
\bibinfo{author}{\bibfnamefont{H.}~\bibnamefont{Romberg}},
  \bibinfo{author}{\bibfnamefont{N.}~\bibnamefont{N{\"u}cker}},
  \bibinfo{author}{\bibfnamefont{J.}~\bibnamefont{Fink}},
  \bibinfo{author}{\bibfnamefont{T.}~\bibnamefont{Wolf}},
  \bibinfo{author}{\bibfnamefont{X.-X.} \bibnamefont{Xi}},
  \bibinfo{author}{\bibfnamefont{B.}~\bibnamefont{Koch}},
  \bibinfo{author}{\bibfnamefont{H.~P.} \bibnamefont{Geserich}},
  \bibinfo{author}{\bibfnamefont{M.}~\bibnamefont{Durrler}},
  \bibinfo{author}{\bibfnamefont{W.}~\bibnamefont{Assmus}}, \bibnamefont{and}
  \bibinfo{author}{\bibfnamefont{B.}~\bibnamefont{Gegenheimer}},
  \bibinfo{journal}{Z. Physik B: Cond. Matter} \textbf{\bibinfo{volume}{78}},
  \bibinfo{pages}{367} (\bibinfo{year}{1990}).

\bibitem[{\citenamefont{Anderson}(1987)}]{Hubbard-HTSC}
\bibinfo{author}{\bibfnamefont{P.~W.} \bibnamefont{Anderson}},
  \bibinfo{journal}{Science} \textbf{\bibinfo{volume}{235}},
  \bibinfo{pages}{1196} (\bibinfo{year}{1987}).

\bibitem[{\citenamefont{Tremblay et~al.}(2006)\citenamefont{Tremblay, Kyung,
  and S{\'e}n{\'e}chal}}]{Hubbard-review}
\bibinfo{author}{\bibfnamefont{A.-M.~S.} \bibnamefont{Tremblay}},
  \bibinfo{author}{\bibfnamefont{B.}~\bibnamefont{Kyung}}, \bibnamefont{and}
  \bibinfo{author}{\bibfnamefont{D.}~\bibnamefont{S{\'e}n{\'e}chal}},
  \bibinfo{journal}{Low Temp. Phys.} \textbf{\bibinfo{volume}{32}},
  \bibinfo{pages}{424} (\bibinfo{year}{2006}).

\bibitem[{\citenamefont{Georges et~al.}(1996)\citenamefont{Georges, Kotliar,
  Krauth, and Rozenberg}}]{georges}
\bibinfo{author}{\bibfnamefont{A.}~\bibnamefont{Georges}},
  \bibinfo{author}{\bibfnamefont{G.}~\bibnamefont{Kotliar}},
  \bibinfo{author}{\bibfnamefont{W.}~\bibnamefont{Krauth}}, \bibnamefont{and}
  \bibinfo{author}{\bibfnamefont{M.~J.} \bibnamefont{Rozenberg}},
  \bibinfo{journal}{Rev. Mod. Phys.} \textbf{\bibinfo{volume}{68}},
  \bibinfo{pages}{13} (\bibinfo{year}{1996}).

\bibitem[{\citenamefont{Civelli et~al.}(2005)\citenamefont{Civelli, Capone,
  Kancharla, Parcollet, and Kotliar}}]{ybco-cdmft}
\bibinfo{author}{\bibfnamefont{M.}~\bibnamefont{Civelli}},
  \bibinfo{author}{\bibfnamefont{M.}~\bibnamefont{Capone}},
  \bibinfo{author}{\bibfnamefont{S.~S.} \bibnamefont{Kancharla}},
  \bibinfo{author}{\bibfnamefont{O.}~\bibnamefont{Parcollet}},
  \bibnamefont{and} \bibinfo{author}{\bibfnamefont{G.}~\bibnamefont{Kotliar}},
  \bibinfo{journal}{Phys. Rev. Lett.} \textbf{\bibinfo{volume}{95}},
  \bibinfo{pages}{106402} (\bibinfo{year}{2005}).

\bibitem[{\citenamefont{Civelli}(2009)}]{htsc-cdmft}
\bibinfo{author}{\bibfnamefont{M.}~\bibnamefont{Civelli}},
  \bibinfo{journal}{Phys. Rev. B} \textbf{\bibinfo{volume}{79}},
  \bibinfo{pages}{195113} (\bibinfo{year}{2009}).

\bibitem[{\citenamefont{Kouba et~al.}(1999)\citenamefont{Kouba, Ambrosch-Draxl,
  and Zangger}}]{ybcolda}
\bibinfo{author}{\bibfnamefont{R.}~\bibnamefont{Kouba}},
  \bibinfo{author}{\bibfnamefont{C.}~\bibnamefont{Ambrosch-Draxl}},
  \bibnamefont{and} \bibinfo{author}{\bibfnamefont{B.}~\bibnamefont{Zangger}},
  \bibinfo{journal}{Phys. Rev. B} \textbf{\bibinfo{volume}{60}},
  \bibinfo{pages}{9321} (\bibinfo{year}{1999}).

\bibitem[{\citenamefont{Khosroabadi et~al.}(2004)\citenamefont{Khosroabadi,
  Mossalla, and Akhavan}}]{ybcopressure}
\bibinfo{author}{\bibfnamefont{H.}~\bibnamefont{Khosroabadi}},
  \bibinfo{author}{\bibfnamefont{B.}~\bibnamefont{Mossalla}}, \bibnamefont{and}
  \bibinfo{author}{\bibfnamefont{M.}~\bibnamefont{Akhavan}},
  \bibinfo{journal}{Phys. Rev. B} \textbf{\bibinfo{volume}{70}},
  \bibinfo{pages}{134509} (\bibinfo{year}{2004}).

\bibitem[{\citenamefont{Runge and Gross}(1984)}]{runge-gross-84}
\bibinfo{author}{\bibfnamefont{E.}~\bibnamefont{Runge}} \bibnamefont{and}
  \bibinfo{author}{\bibfnamefont{E.~K.~U.} \bibnamefont{Gross}},
  \bibinfo{journal}{Phys. Rev. Lett} \textbf{\bibinfo{volume}{52}},
  \bibinfo{pages}{997} (\bibinfo{year}{1984}).

\bibitem[{\citenamefont{Gross and Kohn}(1985)}]{gross-kohn-85}
\bibinfo{author}{\bibfnamefont{E.~K.~U.} \bibnamefont{Gross}} \bibnamefont{and}
  \bibinfo{author}{\bibfnamefont{W.}~\bibnamefont{Kohn}},
  \bibinfo{journal}{Phys. Rev. Lett.} \textbf{\bibinfo{volume}{55}},
  \bibinfo{pages}{2850} (\bibinfo{year}{1985}).

\bibitem[{\citenamefont{Zangwill and Soven}(1980)}]{zangwill-soven}
\bibinfo{author}{\bibfnamefont{A.}~\bibnamefont{Zangwill}} \bibnamefont{and}
  \bibinfo{author}{\bibfnamefont{P.}~\bibnamefont{Soven}},
  \bibinfo{journal}{Phys. Rev. A} \textbf{\bibinfo{volume}{21}},
  \bibinfo{pages}{1561} (\bibinfo{year}{1980}).

\bibitem[{\citenamefont{Olevano et~al.}(1999)\citenamefont{Olevano, Palummo,
  Onida, and {Del Sole}}}]{tddft-eels}
\bibinfo{author}{\bibfnamefont{V.}~\bibnamefont{Olevano}},
  \bibinfo{author}{\bibfnamefont{M.}~\bibnamefont{Palummo}},
  \bibinfo{author}{\bibfnamefont{G.}~\bibnamefont{Onida}}, \bibnamefont{and}
  \bibinfo{author}{\bibfnamefont{R.}~\bibnamefont{{Del Sole}}},
  \bibinfo{journal}{Phys. Rev. B} \textbf{\bibinfo{volume}{60}},
  \bibinfo{pages}{014224} (\bibinfo{year}{1999}).

\bibitem[{\citenamefont{Weissker et~al.}(2006)\citenamefont{Weissker, Serrano,
  Huotari, Bruneval, Sottile, Monaco, Krisch, Olevano, and
  Reining}}]{tddft-ixss}
\bibinfo{author}{\bibfnamefont{H.~C.} \bibnamefont{Weissker}},
  \bibinfo{author}{\bibfnamefont{J.}~\bibnamefont{Serrano}},
  \bibinfo{author}{\bibfnamefont{S.}~\bibnamefont{Huotari}},
  \bibinfo{author}{\bibfnamefont{F.}~\bibnamefont{Bruneval}},
  \bibinfo{author}{\bibfnamefont{F.}~\bibnamefont{Sottile}},
  \bibinfo{author}{\bibfnamefont{G.}~\bibnamefont{Monaco}},
  \bibinfo{author}{\bibfnamefont{M.}~\bibnamefont{Krisch}},
  \bibinfo{author}{\bibfnamefont{V.}~\bibnamefont{Olevano}}, \bibnamefont{and}
  \bibinfo{author}{\bibfnamefont{L.}~\bibnamefont{Reining}},
  \bibinfo{journal}{Phys. Rev. Lett.} \textbf{\bibinfo{volume}{97}},
  \bibinfo{pages}{237602} (\bibinfo{year}{2006}).

\bibitem[{\citenamefont{Verdozzi}(2008)}]{tddftluttinger}
\bibinfo{author}{\bibfnamefont{C.}~\bibnamefont{Verdozzi}},
  \bibinfo{journal}{Phys. Rev. Lett.} \textbf{\bibinfo{volume}{101}},
  \bibinfo{pages}{166401} (\bibinfo{year}{2008}).

\bibitem[{\citenamefont{Li et~al.}(2008)\citenamefont{Li, Xianlong, Kollath,
  and Polini}}]{tddftaldaluttinger}
\bibinfo{author}{\bibfnamefont{W.}~\bibnamefont{Li}},
  \bibinfo{author}{\bibfnamefont{G.}~\bibnamefont{Xianlong}},
  \bibinfo{author}{\bibfnamefont{C.}~\bibnamefont{Kollath}}, \bibnamefont{and}
  \bibinfo{author}{\bibfnamefont{M.}~\bibnamefont{Polini}},
  \bibinfo{journal}{Phys. Rev. B} \textbf{\bibinfo{volume}{78}},
  \bibinfo{pages}{195109} (\bibinfo{year}{2008}).

\bibitem[{\citenamefont{Verbeni et~al.}(2009)\citenamefont{Verbeni,
  Pylkk{\"a}nen, Huotari, Simonelli, Vank{\'o}, Martel, Henriquet, and
  Monaco}}]{verbeni09}
\bibinfo{author}{\bibfnamefont{R.}~\bibnamefont{Verbeni}},
  \bibinfo{author}{\bibfnamefont{T.}~\bibnamefont{Pylkk{\"a}nen}},
  \bibinfo{author}{\bibfnamefont{S.}~\bibnamefont{Huotari}},
  \bibinfo{author}{\bibfnamefont{L.}~\bibnamefont{Simonelli}},
  \bibinfo{author}{\bibfnamefont{G.}~\bibnamefont{Vank{\'o}}},
  \bibinfo{author}{\bibfnamefont{K.}~\bibnamefont{Martel}},
  \bibinfo{author}{\bibfnamefont{C.}~\bibnamefont{Henriquet}},
  \bibnamefont{and} \bibinfo{author}{\bibfnamefont{G.}~\bibnamefont{Monaco}},
  \bibinfo{journal}{J. Synchrotron Radiat.} \textbf{\bibinfo{volume}{16}},
  \bibinfo{pages}{469} (\bibinfo{year}{2009}).

\bibitem[{\citenamefont{Huotari et~al.}(2005)\citenamefont{Huotari, Vank{\'o},
  Albergamo, Ponchut, Graafsma, Henriquet, Verbeni, and Monaco}}]{huotari05}
\bibinfo{author}{\bibfnamefont{S.}~\bibnamefont{Huotari}},
  \bibinfo{author}{\bibfnamefont{G.}~\bibnamefont{Vank{\'o}}},
  \bibinfo{author}{\bibfnamefont{F.}~\bibnamefont{Albergamo}},
  \bibinfo{author}{\bibfnamefont{C.}~\bibnamefont{Ponchut}},
  \bibinfo{author}{\bibfnamefont{H.}~\bibnamefont{Graafsma}},
  \bibinfo{author}{\bibfnamefont{C.}~\bibnamefont{Henriquet}},
  \bibinfo{author}{\bibfnamefont{R.}~\bibnamefont{Verbeni}}, \bibnamefont{and}
  \bibinfo{author}{\bibfnamefont{G.}~\bibnamefont{Monaco}},
  \bibinfo{journal}{J. Synchrotron Radiat.} \textbf{\bibinfo{volume}{12}},
  \bibinfo{pages}{467} (\bibinfo{year}{2005}).

\bibitem[{dp-()}]{dp-web}
\bibinfo{note}{{h}ttp://www.dp-code.org}.

\bibitem[{abi()}]{abinit-web}
\bibinfo{note}{{h}ttp://www.abinit.org}.

\bibitem[{\citenamefont{Gurtubay et~al.}(2005)\citenamefont{Gurtubay, Pitarke,
  Ku, Eguiluz, Larson, Tischler, Zschack, and Finkelstein}}]{gurtubay05}
\bibinfo{author}{\bibfnamefont{I.~G.} \bibnamefont{Gurtubay}},
  \bibinfo{author}{\bibfnamefont{J.~M.} \bibnamefont{Pitarke}},
  \bibinfo{author}{\bibfnamefont{W.}~\bibnamefont{Ku}},
  \bibinfo{author}{\bibfnamefont{A.~G.} \bibnamefont{Eguiluz}},
  \bibinfo{author}{\bibfnamefont{B.~C.} \bibnamefont{Larson}},
  \bibinfo{author}{\bibfnamefont{J.}~\bibnamefont{Tischler}},
  \bibinfo{author}{\bibfnamefont{P.}~\bibnamefont{Zschack}}, \bibnamefont{and}
  \bibinfo{author}{\bibfnamefont{K.~D.} \bibnamefont{Finkelstein}},
  \bibinfo{journal}{Phys. Rev. B} \textbf{\bibinfo{volume}{72}},
  \bibinfo{pages}{125117} (\bibinfo{year}{2005}).

\bibitem[{\citenamefont{Fuggle and M{\aa}rtenson}(1980)}]{fuggle80}
\bibinfo{author}{\bibfnamefont{J.~C.} \bibnamefont{Fuggle}} \bibnamefont{and}
  \bibinfo{author}{\bibfnamefont{N.}~\bibnamefont{M{\aa}rtenson}},
  \bibinfo{journal}{J. El. Spectrosc. Relat. Phenom.}
  \textbf{\bibinfo{volume}{21}}, \bibinfo{pages}{275} (\bibinfo{year}{1980}).

\bibitem[{\citenamefont{Cardona and Ley}(1978)}]{cardona78}
\bibinfo{editor}{\bibfnamefont{M.}~\bibnamefont{Cardona}} \bibnamefont{and}
  \bibinfo{editor}{\bibfnamefont{L.}~\bibnamefont{Ley}}, eds.,
  \emph{\bibinfo{title}{Photoemission in Solids I: General Principles}}
  (\bibinfo{publisher}{Springer-Verlag, Berlin}, \bibinfo{year}{1978}).

\bibitem[{\citenamefont{Pylkk{\"a}nen et~al.}(2010)\citenamefont{Pylkk{\"a}nen,
  Giordano, Chervin, Sakko, Hakala, Soininen, H{\"a}m{\"a}l{\"a}inen, Monaco, and
  Huotari}}]{pylkkanen10}
\bibinfo{author}{\bibfnamefont{T.}~\bibnamefont{Pylkk{\"a}nen}},
  \bibinfo{author}{\bibfnamefont{V.~M.} \bibnamefont{Giordano}},
  \bibinfo{author}{\bibfnamefont{J.~C.} \bibnamefont{Chervin}},
  \bibinfo{author}{\bibfnamefont{A.}~\bibnamefont{Sakko}},
  \bibinfo{author}{\bibfnamefont{M.}~\bibnamefont{Hakala}},
  \bibinfo{author}{\bibfnamefont{J.~A.} \bibnamefont{Soininen}},
  \bibinfo{author}{\bibfnamefont{K.}~\bibnamefont{H{\"a}m{\"a}l{\"a}inen}},
  \bibinfo{author}{\bibfnamefont{G.}~\bibnamefont{Monaco}}, \bibnamefont{and}
  \bibinfo{author}{\bibfnamefont{S.}~\bibnamefont{Huotari}},
  \bibinfo{journal}{J. Phys. Chem B.} \textbf{\bibinfo{volume}{114}},
  \bibinfo{pages}{3804} (\bibinfo{year}{2010}).

\bibitem[{\citenamefont{Lee et~al.}(2008)\citenamefont{Lee, Lin, Cai, Hiraoka,
  Eng, Okuchi, Mao, Meng, Hu, Chow et~al.}}]{lee08}
\bibinfo{author}{\bibfnamefont{S.~K.} \bibnamefont{Lee}},
  \bibinfo{author}{\bibfnamefont{J.-F.} \bibnamefont{Lin}},
  \bibinfo{author}{\bibfnamefont{Y.~Q.} \bibnamefont{Cai}},
  \bibinfo{author}{\bibfnamefont{N.}~\bibnamefont{Hiraoka}},
  \bibinfo{author}{\bibfnamefont{P.~J.} \bibnamefont{Eng}},
  \bibinfo{author}{\bibfnamefont{T.}~\bibnamefont{Okuchi}},
  \bibinfo{author}{\bibfnamefont{H.-K.} \bibnamefont{Mao}},
  \bibinfo{author}{\bibfnamefont{Y.}~\bibnamefont{Meng}},
  \bibinfo{author}{\bibfnamefont{M.~Y.} \bibnamefont{Hu}},
  \bibinfo{author}{\bibfnamefont{P.}~\bibnamefont{Chow}}, \bibnamefont{et~al.},
  \bibinfo{journal}{Proc. Nat. Acad. Sci.} \textbf{\bibinfo{volume}{105}},
  \bibinfo{pages}{7925} (\bibinfo{year}{2008}).

\bibitem[{\citenamefont{Botti et~al.}(2002)\citenamefont{Botti, Vast, Reining,
  Olevano, and Andreani}}]{tddft-anisotropy}
\bibinfo{author}{\bibfnamefont{S.}~\bibnamefont{Botti}},
  \bibinfo{author}{\bibfnamefont{N.}~\bibnamefont{Vast}},
  \bibinfo{author}{\bibfnamefont{L.}~\bibnamefont{Reining}},
  \bibinfo{author}{\bibfnamefont{V.}~\bibnamefont{Olevano}}, \bibnamefont{and}
  \bibinfo{author}{\bibfnamefont{L.~C.} \bibnamefont{Andreani}},
  \bibinfo{journal}{Phys. Rev. Lett.} \textbf{\bibinfo{volume}{89}},
  \bibinfo{pages}{216803} (\bibinfo{year}{2002}).

\bibitem[{\citenamefont{Sottile et~al.}(2005)\citenamefont{Sottile, Bruneval,
  Marinopoulos, Dash, Botti, Olevano, Vast, Rubio, and
  Reining}}]{tddft-longrange}
\bibinfo{author}{\bibfnamefont{F.}~\bibnamefont{Sottile}},
  \bibinfo{author}{\bibfnamefont{F.}~\bibnamefont{Bruneval}},
  \bibinfo{author}{\bibfnamefont{A.~G.} \bibnamefont{Marinopoulos}},
  \bibinfo{author}{\bibfnamefont{L.~K.} \bibnamefont{Dash}},
  \bibinfo{author}{\bibfnamefont{S.}~\bibnamefont{Botti}},
  \bibinfo{author}{\bibfnamefont{V.}~\bibnamefont{Olevano}},
  \bibinfo{author}{\bibfnamefont{N.}~\bibnamefont{Vast}},
  \bibinfo{author}{\bibfnamefont{A.}~\bibnamefont{Rubio}}, \bibnamefont{and}
  \bibinfo{author}{\bibfnamefont{L.}~\bibnamefont{Reining}},
  \bibinfo{journal}{Int. J. Quantum Chem.} \textbf{\bibinfo{volume}{102}},
  \bibinfo{pages}{684} (\bibinfo{year}{2005}).

\bibitem[{\citenamefont{{Del Sole} et~al.}(2003)\citenamefont{{Del Sole},
  Adragna, Olevano, and Reining}}]{tddft-dynamical}
\bibinfo{author}{\bibfnamefont{R.}~\bibnamefont{{Del Sole}}},
  \bibinfo{author}{\bibfnamefont{G.}~\bibnamefont{Adragna}},
  \bibinfo{author}{\bibfnamefont{V.}~\bibnamefont{Olevano}}, \bibnamefont{and}
  \bibinfo{author}{\bibfnamefont{L.}~\bibnamefont{Reining}},
  \bibinfo{journal}{Phys. Rev. B} \textbf{\bibinfo{volume}{67}},
  \bibinfo{pages}{045207} (\bibinfo{year}{2003}).

\bibitem[{\citenamefont{Rignanese et~al.}(2001)\citenamefont{Rignanese, Blase,
  and Louie}}]{quasipart-effects}
\bibinfo{author}{\bibfnamefont{G.-M.} \bibnamefont{Rignanese}},
  \bibinfo{author}{\bibfnamefont{X.}~\bibnamefont{Blase}}, \bibnamefont{and}
  \bibinfo{author}{\bibfnamefont{S.~G.} \bibnamefont{Louie}},
  \bibinfo{journal}{Phys. Rev. Lett.} \textbf{\bibinfo{volume}{86}},
  \bibinfo{pages}{2110} (\bibinfo{year}{2001}).

\bibitem[{\citenamefont{Albrecht et~al.}(1998)\citenamefont{Albrecht, Reining,
  {Del Sole}, and Onida}}]{excitonic-effects}
\bibinfo{author}{\bibfnamefont{S.}~\bibnamefont{Albrecht}},
  \bibinfo{author}{\bibfnamefont{L.}~\bibnamefont{Reining}},
  \bibinfo{author}{\bibfnamefont{R.}~\bibnamefont{{Del Sole}}},
  \bibnamefont{and} \bibinfo{author}{\bibfnamefont{G.}~\bibnamefont{Onida}},
  \bibinfo{journal}{Phys. Rev. Lett.} \textbf{\bibinfo{volume}{80}},
  \bibinfo{pages}{4510} (\bibinfo{year}{1998}).

\bibitem[{\citenamefont{Albrecht et~al.}(1999)\citenamefont{Albrecht, Reining,
  Onida, Olevano, and {Del Sole}}}]{excitonic-effects-rep}
\bibinfo{author}{\bibfnamefont{S.}~\bibnamefont{Albrecht}},
  \bibinfo{author}{\bibfnamefont{L.}~\bibnamefont{Reining}},
  \bibinfo{author}{\bibfnamefont{G.}~\bibnamefont{Onida}},
  \bibinfo{author}{\bibfnamefont{V.}~\bibnamefont{Olevano}}, \bibnamefont{and}
  \bibinfo{author}{\bibfnamefont{R.}~\bibnamefont{{Del Sole}}},
  \bibinfo{journal}{Phys. Rev. Lett.} \textbf{\bibinfo{volume}{83}},
  \bibinfo{pages}{3971} (\bibinfo{year}{1999}).

\bibitem[{\citenamefont{Galambosi et~al.}(2005)\citenamefont{Galambosi,
  Soininen, Mattila, Huotari, Manninen, Vank{\'o}, Zhigadlo, Karpinski, and
  H{\"a}m{\"a}l{\"a}inen}}]{galambosi05}
\bibinfo{author}{\bibfnamefont{S.}~\bibnamefont{Galambosi}},
  \bibinfo{author}{\bibfnamefont{J.~A.} \bibnamefont{Soininen}},
  \bibinfo{author}{\bibfnamefont{A.}~\bibnamefont{Mattila}},
  \bibinfo{author}{\bibfnamefont{S.}~\bibnamefont{Huotari}},
  \bibinfo{author}{\bibfnamefont{S.}~\bibnamefont{Manninen}},
  \bibinfo{author}{\bibfnamefont{G.}~\bibnamefont{Vank{\'o}}},
  \bibinfo{author}{\bibfnamefont{N.~D.} \bibnamefont{Zhigadlo}},
  \bibinfo{author}{\bibfnamefont{J.}~\bibnamefont{Karpinski}},
  \bibnamefont{and}
  \bibinfo{author}{\bibfnamefont{K.}~\bibnamefont{H{\"a}m{\"a}l{\"a}inen}},
  \bibinfo{journal}{Phys. Rev. B} \textbf{\bibinfo{volume}{71}},
  \bibinfo{pages}{060504(R)} (\bibinfo{year}{2005}).

\bibitem[{\citenamefont{Huotari et~al.}(2009)\citenamefont{Huotari, Sternemann,
  Troparevsky, Eguiluz, Volmer, Sternemann, M{\"u}ller, Monaco, and
  Sch{\"u}lke}}]{huotari09}
\bibinfo{author}{\bibfnamefont{S.}~\bibnamefont{Huotari}},
  \bibinfo{author}{\bibfnamefont{C.}~\bibnamefont{Sternemann}},
  \bibinfo{author}{\bibfnamefont{M.~C.} \bibnamefont{Troparevsky}},
  \bibinfo{author}{\bibfnamefont{A.~G.} \bibnamefont{Eguiluz}},
  \bibinfo{author}{\bibfnamefont{M.}~\bibnamefont{Volmer}},
  \bibinfo{author}{\bibfnamefont{H.}~\bibnamefont{Sternemann}},
  \bibinfo{author}{\bibfnamefont{H.}~\bibnamefont{M{\"u}ller}},
  \bibinfo{author}{\bibfnamefont{G.}~\bibnamefont{Monaco}}, \bibnamefont{and}
  \bibinfo{author}{\bibfnamefont{W.}~\bibnamefont{Sch{\"u}lke}},
  \bibinfo{journal}{Phys. Rev. B} \textbf{\bibinfo{volume}{80}},
  \bibinfo{pages}{155107} (\bibinfo{year}{2009}).

\bibitem[{\citenamefont{Sturm and Oliveira}(1981)}]{sturm81}
\bibinfo{author}{\bibfnamefont{K.}~\bibnamefont{Sturm}} \bibnamefont{and}
  \bibinfo{author}{\bibfnamefont{L.~E.} \bibnamefont{Oliveira}},
  \bibinfo{journal}{Phys. Rev. B} \textbf{\bibinfo{volume}{24}},
  \bibinfo{pages}{3054} (\bibinfo{year}{1981}).

\bibitem[{\citenamefont{Eguiluz et~al.}(2000)\citenamefont{Eguiluz, Ku, and
  Sullivan}}]{eguiluz00}
\bibinfo{author}{\bibfnamefont{A.~G.} \bibnamefont{Eguiluz}},
  \bibinfo{author}{\bibfnamefont{W.}~\bibnamefont{Ku}}, \bibnamefont{and}
  \bibinfo{author}{\bibfnamefont{J.~M.} \bibnamefont{Sullivan}},
  \bibinfo{journal}{J. Phys. Chem. Solids} \textbf{\bibinfo{volume}{61}},
  \bibinfo{pages}{383} (\bibinfo{year}{2000}).

\end{thebibliography}

\end{document}